# L'interphase dans les modèles de prédiction du comportement mécanique pour les nanocomposites

*Interphase in the mechanical behaviour prediction models for nanocomposites*

**Nicolas Raoux[1,2], Abdelkibir Benelfellah[1,2], Nourredine Aït Hocine[2]**

[1] : Direction de la Recherche et de l'Innovation à l'IPSA (DR2I), IPSA
63 boulevard de Brandebourg, 94200 Ivry-sur-Seine
e-mail : nicolas1.raoux@ipsa.fr et abdelkibir.benelfellah@ipsa.fr

[2] : INSA CVL, Univ. Tours, Univ. Orléans, LaMé
3 Rue de la Chocolaterie - CS 23410, 41034 Blois cedex, France.
e-mail : nourredine.aithocine@insa-cvl.fr

## Résumé

Les polymères sont de plus en plus utilisés dans le secteur du transport du fait de leurs nombreux avantages ; légèreté, résistance à la corrosion, facilité d'usinage... Cependant, la plupart présentent des propriétés mécaniques limitées. Afin d'améliorer ces dernières, une des solutions est l'ajout de nanorenforts, ce type de matériau est qualifié de nanocomposite. La particularité de ces nanocomposites vient de l'influence de l'interphase, qui est une zone d'interaction entre les nanocharges et la matrice. Il est donc impératif de la prendre en considération dans les modèles de prédiction mécanique. Plusieurs schémas analytiques permettent d'étudier les nanocomposites. Certains modèles utilisent une représentation en série ou en parallèle des phases (loi des mélanges, modèle de Ji ...), tandis que d'autres reposent sur l'homogénéisation par champs moyens (Eshelby, Auto-cohérent, Mori-Tanaka, Double Inclusion ...).
Cependant, observer directement l'interphase et mesurer ses propriétés expérimentalement s'avère complexe, du fait de sa taille nanométrique. Par conséquent, il est laborieux de vérifier la cohérence de sa prise en compte dans les schémas analytiques. Afin de comparer ces schémas, certains cas limites de l'interphase sont étudiés. Un modèle numérique par champs complets est également exploité comme référence pour deux nanocomposites avec des interphases distinctes.

## Abstract

Polymers are increasingly used in the transport sector due to their many advantages; lightness, corrosion resistance, ease to process... However, most of them have limited mechanical properties. To improve these latter, one of the solutions is the addition of nano-reinforcements, this type of material is called nanocomposite. The particularity of these nanocomposites comes from the influence of the interphase, which is an interaction zone between the nanofillers and the matrix. It is therefore imperative to take it into account in the mechanical prediction models.
Several analytical schemes are available to study nanocomposites. Some models use a serial or parallel representation of the phases (rule of mixture, Ji model, ...), while others rely on the mean field homogenization approach (Eshelby, Self-consistent, Mori-Tanaka, Double Inclusion, ...).
However, directly observing the interphase and its properties experimentally is complex due to its size. Consequently, it is laborious to check the consistency of its consideration in analytical schemes. To compare these schemes, some limit cases of the interphase are studied. A numerical model is also used as a reference for two nanocomposites with distinct interphases.

**Mots Clés :** Nanocomposite, interphase, propriétés mécaniques, schéma de prédiction, modélisation numérique

## 1 Introduction

Les polymères présentent multiples avantages tels que leur légèreté, capacité d'isolation thermique et électrique, propriétés optiques, résistance à la corrosion, facilité de traitement et faible coût d'usinage. De ce fait, ils sont largement utilisés dans l'automobile, le textile, la construction... Cependant, ils exhibent de faibles propriétés mécaniques.

L'ajout de renforts dans une matrice polymère permet d'améliorer leurs propriétés mécaniques. Les dimensions des inclusions dans les nanocomposites sont de l'ordre du nanomètre. L'interphase désigne la zone de matrice entourant l'inclusion dont les propriétés sont affectées par l'interaction entre les constituants. En raison de sa taille de l'ordre du nanomètre, son influence est significative principalement dans les nanocomposites. C'est ce qui permet à ces matériaux d'offrir une amélioration notable des propriétés même à de faibles taux d'inclusion. L'interphase peut améliorer différents types de propriétés, on se contentera d'étudier son impact sur les propriétés mécaniques.

Les modèles de prédiction des propriétés mécaniques effectives sont essentiels pour l'étude des structures nanocomposites. L'approche numérique permet une homogénéisation par champs complets, en étudiant un volume élémentaire représentatif (VER) à l'aide d'un logiciel de calcul par éléments finis (FEM). Il existe également des approches analytiques d'homogénéisation. Parmi celles-ci, une première catégorie considère les phases disposées en parallèle ou en série (loi des mélanges (RoM), loi des mélanges inverse (iRoM), schéma de Ji [1]…). Une autre approche consiste à définir une formulation analytique simple avec un paramètre empirique (Halpin-Tsai [2]). Finalement, de nombreux schémas fréquemment employés reposent sur l'homogénéisation par champs moyens (MFH) (auto-cohérent, Mori-Tanaka (MT) [3] …), hypothèse introduite par Eshelby [4]. En particulier, pour les nanocomposites le schéma de double inclusion (DI) [5] ressort régulièrement.

La prise en compte de l'interphase est primordiale pour l'étude des nanocomposites. Cependant, les méthodes permettant de l'observer directement sont limitées. Expérimentalement, il est possible d'utiliser un microscope à force atomique (AFM) et numériquement, des simulations de dynamique moléculaire (MD). Mais, pour une interphase de quelques nanomètres d'épaisseur, ces procédés sont complexes à mettre en place et les valeurs obtenues souvent imprécises. Par conséquent, les propriétés de l'interphase sont communément déterminées de manière indirecte, ce qui rend la comparaison des modèles employés difficile.

Cet article propose une comparaison des différents modèles et en particulier leur considération de l'interphase. Pour se faire, des cas limites de l'interphase sont introduits afin de vérifier leur cohérence. Par la suite, comme les modèles numériques ne nécessitent pas d'hypothèses supplémentaires pour considérer la disposition de l'interphase, ils servent de référence dans deux cas d'études avec différentes interphases.

## 2 Modèles de prédiction du comportement élastique

Les matériaux nanocomposites présentent en général un comportement complexe. Pour se focaliser sur les propriétés élastiques effectives du matériau nanocomposite et s'affranchir des autres mécanismes de comportement (plasticité, viscosité, endommagement …), on considère que les contraintes restent inférieures à la limite élastique et que la vitesse de sollicitation est quasi statique. Différentes approches numériques et analytiques seront étudiées pour déterminer la rigidité homogénéisée du nanocomposite. Certaines, qui sont régulièrement exploitées dans la littérature, sont présentées dans cet article.

### 2.1 Modèle numérique

En considérant l'hypothèse de périodicité et de continuité du VER, l'approche numérique procure des résultats précis, mais elle est également très coûteuse en termes de temps de calcul. À l'aide d'un logiciel de calcul par éléments finis (FEM), cette méthode permet d'obtenir les propriétés élastiques homogènes en considérant les champs complets. Pour se faire, il est nécessaire de rentrer les propriétés de chaque phase ainsi que de déterminer le VER, qui est un volume infinitésimal se répétant pour former la structure. Dans un même temps, ce volume doit être suffisamment grand pour correctement prendre en compte la distribution des renforts. Ce deuxième point est déterminé grâce à une étude de stabilité. Un VER est considéré comme stable lorsque la variation des résultats obtenus entre deux études est négligeable. Pour atteindre cet objectif, il est possible de changer la taille du volume (soit le nombre de renforts considérés) et le nombre de réalisations [6].

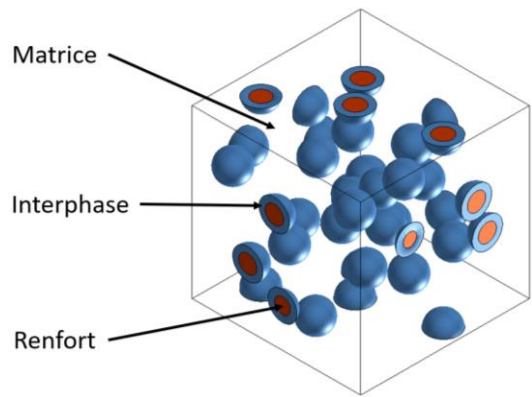

*Fig. 1. VER d'un nanocomposite avec 30 renforts sphériques répartis de manière aléatoire*

Un exemple de VER est donné (Fig. 1) pour 30 renforts sphériques (en orange), entourés d'interphase (bleu), distribués de manière aléatoire dans un cube de matrice. Le caractère aléatoire de la distribution rend chaque volume généré différent. Ainsi, la moyenne sur plusieurs volumes est considérée, ce nombre de volumes est le nombre de réalisations.

Il a été prouvé et admis que le calcul par FEM donne des résultats consistants [7], pour les microcomposites entre autres, tant qu'un maillage suffisamment fin est utilisé. De plus, aucune nouvelle hypothèse simplificatrice n'a été introduite pour l'étude des nanocomposites. Il est donc possible d'assumer que les estimations numériques conservent leur cohérence. Cependant, la définition et l'étude du VER sont coûteuses en temps de calcul. De cette perspective, les schémas analytiques sont plus avantageux et permettent d'isoler plus facilement l'influence d'un paramètre en particulier. Néanmoins, ils nécessitent davantage d'hypothèses simplificatrices. Ainsi, il existe de nombreux modèles analytiques, reposant sur différentes hypothèses. Certains de ceux implémentés pour l'étude des nanocomposites sont introduits dans les sections suivantes.

**2.2 Représentation en parallèle et en série**

Une première classe de modèles repose sur la représentation en parallèle ou en série des phases, respectivement appelées RoM et iRoM. Une approche un peu plus robuste, qui considère de manière simplifiée la forme des phases, est le schéma de Ji [1]. Il s'agit d'un mélange des deux approches précédentes.

Ce schéma permet d'obtenir une expression du module de Young homogénéisé du composite ($E_c$) (Eq. 1). Ce module dépend des modules de Young pour les différentes phases $i$ ($E_i$). Les indices $m$, $I$, $r$ et $c$ correspondent respectivement à la matrice, l'interphase, les renforts et le composite homogénéisé. De plus, dans le schéma de Ji l'ensemble des phases sont représentées sur un plan, dans un carré de côté normé (égale à 1). Dans le cas où le renfort et l'interphase sont représentés par des carrés, alors $\lambda$ (Eq. 2) est la taille du côté du carré du renfort et $\alpha$ celui de l'ensemble interphase/renfort (pour des renforts sphériques (Eq. 3)). Le terme $v_i$ correspond à la fraction volumique de la phase $i$, $r$ est le rayon de l'inclusion et $t_I$ est l'épaisseur de l'interphase.

$$\frac{E_c}{E_m} = \left[(1-\alpha) + \frac{\alpha - \lambda}{(1-\alpha) + \alpha \frac{E_I}{E_m}} + \frac{\lambda}{(1-\alpha) + (\alpha - \lambda)\frac{E_I}{E_m} + \lambda \frac{E_r}{E_m}}\right]^{-1} \quad \text{(Eq. 1)}$$

$$\lambda = \sqrt{v_r} \quad \text{(Eq. 2)}$$

$$\alpha = \sqrt{v_r \left(\frac{r + t_I}{r}\right)^3} \quad \text{(Eq. 3)}$$

Ce schéma permet d'obtenir une première estimation avec une formulation simple et une considération de la géométrie. Malheureusement, les hypothèses effectuées sur la répartition du chargement en série et en parallèle sont relativement contraignantes, de plus la considération de la forme est limitée dans ce modèle.

**2.3 Schéma semi-empirique**

Il existe d'autres approches pour traiter la forme de l'inclusion. Entre autres, le schéma de Halpin-Tsai [2] propose une formulation analytique (Eq. 4) avec un paramètre empirique, $\xi_i$, pour la forme de l'inclusion. Où $\eta_i$ (Eq. 5) dépend des modules et de $\xi_i$.

$$\frac{E_c}{E_m} = \frac{1 + v_r \eta_r \xi_r + v_I \eta_I \xi_I}{1 - v_r \eta_r - v_I \eta_I} \quad \text{(Eq. 4)}$$

$$\eta_i = \frac{\frac{E_i}{E_m} - 1}{\frac{E_i}{E_m} + \xi_i} \quad \text{(Eq. 5)}$$

Le modèle de Halpin-Tsai est souvent présenté comme un schéma semi-empirique du fait qu'il dépend en partie des paramètres analytiques (module et fraction volumique), auxquels vient s'ajouter un paramètre empirique. Sa formulation simple est son avantage majeur, mais l'introduction d'un paramètre empirique implique la nécessité d'une base de données servant de référence, de plus la valeur de ce terme n'est pas justifiable théoriquement.

**2.4 Schémas d'homogénéisation par champs moyens (MFH)**

Basés sur l'hypothèse d'Eshelby [4], les schémas MFH sont régulièrement employés pour les microcomposites. Ils considèrent que les variables (contraintes, déformations…) admettent une valeur moyenne et uniforme dans chaque phase. À partir de cette hypothèse et de la loi de Hooke en 3D (Eq. 6), une formulation pour le tenseur de rigidité équivalent est obtenue (Eq. 7). Les contraintes ($\boldsymbol{\sigma}$) et les déformations ($\boldsymbol{\epsilon}$) s'expriment comme des tenseurs d'ordre 2, tandis que le tenseur de rigidité ($\boldsymbol{C}$) et de localisation des déformations ($\boldsymbol{A}$) sont d'ordre 4.

$$\boldsymbol{\sigma} = \boldsymbol{C} : \boldsymbol{\epsilon} \quad \text{(Eq. 6)}$$

$$\boldsymbol{C}_c = \sum_i v_i \boldsymbol{C}_i : \boldsymbol{A}_i \quad \text{(Eq. 7)}$$

Il existe de nombreux schémas basés sur ces hypothèses. Les différences dans les hypothèses se retrouvent dans l'expression du tenseur de localisation. Parmi ces schémas, le schéma de MT [3] est fréquemment utilisé et prodigue des prédictions fidèles de la rigidité. Il considère l'interaction entre les différents renforts, ce qui lui permet d'obtenir des valeurs proches des résultats expérimentaux même lorsque la fraction volumique augmente. Son tenseur de localisation ($A_i^{MT}$) (Eq. 8) dépend du tenseur de localisation du modèle d'Eshelby ($A_i^{Es}$) pour la phase $i$ correspondante (Eq. 9) avec le tenseur d'Eshelby ($S_i^E$) qui dépend de la forme de la phase.

$$A_i^{MT} = T_i : \left(\sum_s v_s T_s\right)^{-1} \quad \text{(Eq. 8)}$$

$$T_i = A_i^{Es} = \left(I + S_i^E : C_m^{-1} : (C_i - C_m)\right)^{-1} \quad \text{(Eq. 9)}$$

Ce schéma peut être appliqué en considérant multiple phases $i$, pour les nanocomposites 3 phases sont essentielles à différencier (matrice, renfort et interphase) (MT3). Cependant, il n'inclut aucune interaction directe entre le renfort et l'interphase. Ainsi, l'influence de la morphologie particulière des nanocomposites, où l'interphase englobe les renforts, est négligée.

Le schéma DI, développé par Hori et Nemat-Nasser [5], tente de prendre en compte cet arrangement par superposition. Cependant, le seul terme qui est ajouté dans ce modèle dépend de la différence entre le tenseur de forme de l'interphase et du renfort. Donc, dans le cas où le facteur d'aspect est identique pour les deux phases, ce schéma est équivalant à MT3 [8]. De manière générale l'écart entre les tenseurs de forme du renfort et de l'interphase est minime, voire nul dans le cas sphérique, et il est souvent négligé dans la littérature [9]. Donc la considération de l'interphase semble encore limitée.

**2.5 Homogénéisation incrémentale**

Afin de prendre en compte au mieux cette morphologie des nanocomposites, certaines études ont développé des schémas incrémentaux qui procèdent à l'homogénéisation en plusieurs étapes. Ces schémas reposent sur une ou plusieurs des approches évoquées précédemment.

Un de ces schémas consiste à considérer d'abord la matrice avec l'interphase comme un milieu équivalent, puis d'y intégrer le renfort [10]. Il est également possible dans cette approche de subdiviser l'interphase en différentes couches. Chaque étape de l'homogénéisation est effectuée en utilisant le schéma de MT. Cette méthode sera notée comme MT-inc.

Une approche réciproque consiste à considérer renfort et interphase comme une seule particule équivalente. Différentes méthodes peuvent être employées pour obtenir les propriétés de cette particule équivalente. Par exemple, en utilisant iRoM [6,11], le schéma de MT ou encore un résultat FEM [12]. Dans chaque cas, la deuxième étape, qui est l'incorporation de cette particule équivalente dans la matrice, est réalisée avec MT. Les modèles vont être désignés comme MT-eq pour le cas où la première étape est effectuée avec iRoM et MT$^2$ si MT est employé aux deux étapes.

Chaque modèle présente des limitations et perd en précision en fonction de ses hypothèses simplificatrices. Les schémas vont être comparés pour certains cas concrets, afin d'estimer ceux semblant le mieux adaptés pour prédire le comportement des nanocomposites avec l'interphase.

**3   Comparaison des différents modèles**

Afin d'effectuer une comparaison sur une base commune, il est nécessaire de conserver les mêmes paramètres dans chaque modèle. En particulier, pour évaluer la considération de l'interphase, ses propriétés doivent être définies au préalable. Malheureusement, les études observées ne procurent pas ces données de manière directe, ce qui les rend obsolètes pour la comparaison. Par conséquent, la cohérence des modèles est examinée à l'aide de cas limites de l'interphase. Par la suite, un modèle numérique est pris comme référence pour deux interphases différentes.

### 3.1 Comparaison des modèles pour des cas limites de l'interphase

Dans un premier temps, des cas particuliers de l'interphase sont considérés, pour lesquels la structure est équivalente à un matériau à 2 phases. Dans ces circonstances, il a été prouvé que les modèles numériques ou le schéma de MT procurent des estimations proches de la réalité, il est donc possible de les utiliser comme référence. Pour tous les cas étudiés, la matrice a un module $E_m = 2,75\ MPa$, les inclusions sont sphériques de rayon $15\ nm$ et de module $E_r = 88,7\ MPa$. Chacun des cas présente des résultats fondamentalement différents en fonction du module assigné à l'interphase, tandis que son épaisseur n'influence pas les tendances observées.

Dans le premier cas, l'interphase est assimilée à du vide avec une rigidité nulle ($E_I = 0\ MPa$). Aucun transfert de charge n'est engendré entre la matrice et le renfort. L'ensemble interphase plus renfort peut donc être modélisé comme une seule inclusion de vide. Le deuxième cas traite l'interphase comme une phase de matrice ($E_I = E_m$). Le modèle est donc équivalant à un matériau à 2 phases sans considération de l'interphase. Le troisième cas est celui où l'interphase agit comme une prolongation du renfort ($E_I = E_r$). Cela peut être représenté comme une inclusion (interphase plus renfort), avec les propriétés du renfort, noyée dans la phase matrice. Le quatrième cas observe une interphase infiniment rigide ($E_I \to +\infty$). La déformation de l'interphase est négligeable et il en va de même pour le renfort à l'intérieur. Le module du renfort n'exerce ainsi aucune influence. Donc l'inclusion équivalente, du renfort plus de l'interphase, peut être considérée comme étant infiniment rigide.

| Schémas : | $E_I = 0$ | $E_I = E_m$ | $E_I = E_r$ | $E_I \to +\infty$ |
|---|---|---|---|---|
| **RoM** | $--$ | $-$ | $-$ | $--$ |
| **Ji** | $--$ | $0$ | $+$ | $-$ |
| **Halpin-Tsaï** | $--$ | $+$ | $0$ | $--$ |
| **MT 2 phases** | $--$ | $++$ | $-$ | $--$ |
| **DI (=MT3)** | $--$ | $++$ | $++$ | $--$ |
| **MT-inc** | $--$ | $++$ | $+$ | $--$ |
| **MT-eq** | $++$ | $0$ | $++$ | $--$ |
| **MT$^2$** | $++$ | $++$ | $++$ | $++$ |

*Tab. 1. Tableau de comparaison des différents schémas analytiques pour les cas limites de l'interphase*

Les différents schémas sont étudiés pour les 4 cas particuliers (Tab. 1). La notation va de « $--$ » à « $++$ », allant respectivement d'une tendance opposée à des résultats identiques par rapport à la référence à 2 phases. La plupart des schémas suivent la tendance pour les cas 2 et 3. En revanche, pour les cas extrêmes (1 et 4), la majorité d'entre eux divergent. MT-eq est cohérent pour le cas 1, mais pas dans les cas 2 et 4. Il accorde trop d'importance à la phase la plus souple entre le renfort et l'interphase. Finalement, seul MT$^2$ présente des résultats consistants pour tous les cas considérés. Effectivement, ce schéma semble être le seul à accorder une importance suffisante à l'impact de l'interphase par son transfert de charge au renfort du fait de sa morphologie.

### 3.2 Comparaison avec le modèle numérique

Dans un second temps, en plus des valeurs particulières de l'interphase, deux situations supplémentaires sont considérées. Les propriétés de la matrice et des renforts restent identiques. L'interphase prend une épaisseur de $10\ nm$ tandis que deux cas vont être considérés pour son module ; le cas souple ($E_I = 0{,}5E_m$) et le cas rigide ($E_I = 5E_m$). Seuls les modèles semblant les plus robustes dans la comparaison précédente sont étudiés (DI, MT-eq et MT$^2$). Un modèle numérique (FEM) sert de référence comme il ne comprend pas d'hypothèse simplificatrice supplémentaire comparée aux études à 2 phases. Effectivement, la morphologie particulière des nanocomposites est directement reproduite dans la géométrie de la représentation numérique. Pour l'étude FEM, un RVE contenant 30 particules avec 3 réalisations est utilisé. Les résultats sont comparés pour une fraction volumique de renforts variant entre 0 et 5% (Fig. 2).

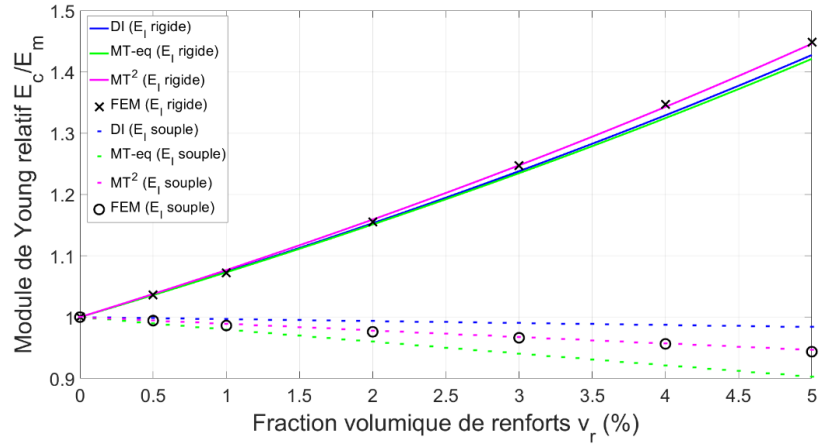

*Fig. 2. Graphique de comparaison du module relatif donné par les schémas entre 0 et 5% pour une interphase souple ou rigide*

Pour le cas rigide, tous les schémas considérés présentent une estimation adéquate, ce qui est en accord avec les résultats de la littérature. En revanche, pour le cas souple des différences sont observées. DI est trop rigide, car il ne prend pas en compte la perte de transfert de charges par l'interphase. Au contraire, MT-eq est trop souple, trop d'importance est accordée à l'interphase dans la particule équivalente. Finalement, seul le schéma MT$^2$ semble procurer une bonne prédiction du module de Young. Cela est rendu possible par ce modèle grâce à sa considération adéquate du renfort totalement englobé par l'interphase, ainsi les chargements subis par le renfort sont ceux transmis par cette dernière.

## 4 Conclusion

Les matériaux nanocomposites présentent une structure similaire aux microcomposites. Une des différences notables reste la prise en compte de l'effet de l'interphase lors de la modélisation. Afin de prédire leur module de Young homogène, il est donc possible d'utiliser les schémas micromécaniques, dont la robustesse a déjà été approuvée, en y ajoutant cette phase supplémentaire. Pour correctement prendre en compte qu'elle englobe les renforts, de nouvelles hypothèses sont posées. Les hypothèses principales sur lesquelles les schémas ont été construits sont présentées au début de l'article.

Par la suite, afin de comparer ces nouveaux schémas des cas limites de l'interphase ont été introduits. Les cas où l'interphase est assimilée à la matrice ou aux renforts, présentent des résultats cohérents pour la plupart des schémas, plus ou moins précis selon la rigueur du modèle d'origine. Cependant, pour les cas extrêmes, interphase agissant comme du vide ou infiniment rigide, aucun des modèles d'homogénéisation en une étape ne parvient à procurer la tendance attendue. Une sous-estimation de l'influence de l'interphase est observée, en particulier dans son transfert de charge aux renforts. Les seuls schémas qui parviennent à prendre cet aspect en compte sont les schémas en deux étapes qui commencent par calculer une particule équivalente. Finalement, seul le schéma MT$^2$ procure des prédictions satisfaisantes pour l'ensemble des conditions limites introduites.

Dans un deuxième temps, les schémas les plus robustes sont comparés à un modèle numérique servant de référence. Dans cette étude, pour une interphase rigide tous les modèles offrent des résultats proches de ceux de références. En revanche dans le cas souple, où il y a une perte du transfert des contraintes due à l'interphase, MT$^2$ ressort à nouveau comme le modèle le plus cohérent.

Ces deux études tendent à montrer que parmi tous les modèles introduits le plus robuste semble être le schéma en deux étapes MT$^2$. Cependant, il a été observé dans la littérature que MT procure des résultats satisfaisants lorsque la fraction volumique de renforts est inférieure ou égale à 30%. Or, lors de la première étape du calcul de la particule équivalente, la proportion du renfort dans l'interphase est en général supérieure à ce seuil. Dans cette étude, cela n'a pas semblé générer d'incohérence dans les résultats, mais à l'avenir il est possible que la précision du schéma en soit impactée dans certains cas.

**Références**